\documentclass[a4paper,fleqn]{cas-dc}

\usepackage[numbers]{natbib}
\usepackage{times}
\usepackage{soul}
\usepackage{url}
\usepackage[utf8]{inputenc}
\usepackage[small]{caption}
\usepackage{amsmath}
\usepackage{amsthm}
\usepackage{booktabs}
\usepackage{xcolor}
\usepackage{tabularx}
\usepackage{xspace}
\usepackage{listings}
\usepackage{subfigure}
\usepackage{pythonhighlight}
\usepackage{amsfonts}
\usepackage{amssymb}
\usepackage{multirow}
\usepackage{multicol}
\usepackage[frozencache,cachedir=.]{minted}
\usepackage{tcolorbox}
\tcbuselibrary{minted,breakable,xparse,skins}

\definecolor{bg}{gray}{1.0}
\DeclareTCBListing{mintedbox}{O{}m!O{}}{%
  breakable=true,
  listing engine=minted,
  listing only,
  minted language=#2,
  minted style=default,
  minted options={%
    linenos,
    gobble=0,
    breaklines=true,
    breakafter=,,
    fontsize=\normalsize,
    numbersep=8pt,
    #1},
  boxsep=0pt,
  left skip=0pt,
  right skip=0pt,
  left=15pt,
  right=0pt,
  top=3pt,
  bottom=2pt,
  arc=5pt,
  leftrule=0pt,
  rightrule=0pt,
  bottomrule=0.5pt,
  toprule=0.5pt,
  colback=bg,
  colframe=black!70,
  enhanced,
  overlay={%
    \begin{tcbclipinterior}
    \fill[black!20!white] (frame.south west) rectangle ([xshift=10pt]frame.north west);
    \end{tcbclipinterior}},
  #3}

\newcommand{\tool}{QueCos\xspace}

\def\tsc#1{\csdef{#1}{\textsc{\lowercase{#1}}\xspace}}
\tsc{WGM}
\tsc{QE}
\tsc{EP}
\tsc{PMS}
\tsc{BEC}
\tsc{DE}

\begin{document}
\let\WriteBookmarks\relax
\def\floatpagepagefraction{1}
\def\textpagefraction{.001}
\shorttitle{QueCos}
\shortauthors{Wang et~al.}

\title [mode = title]{Enriching Query Semantics for Code Search with Reinforcement Learning}                      



\author[1]{Chaozheng Wang}
\ead{wangchaozheng@stu.hit.edu.cn}

\author[1]{Zhenhao Nong}
\ead{nongzhenhao@stu.hit.edu.cn}

\author[1]{Cuiyun Gao}
\ead{gaocuiyun@hit.edu.cn}
\cormark[1]

\author[1]{Zongjie Li}
\ead{lizongjie@stu.hit.edu.cn}

\author[2]{Jichuan Zeng}
\ead{jczeng@cse.cuhk.edu.hk}

\author[3]{Zhenchang Xing}
\ead{zhenchang.xing@anu.edu.au}

\author[4]{Yang Liu}
\ead{yangliu@ntu.edu.sg}

\address[1]{School of Computer Science and Technology, Harbin Institute of Technology, Shenzhen, China}

\address[2]{Department of Computer Science and Engineering, The Chinese University of Hong Kong, Hong Kong, China}

\address[3]{Research School of Computer Science, Australian National University, Australia}

\address[4]{School of Computer Science and Engineering, Nanyang Technology University, Singapore}








\cortext[cor1]{Corresponding author}

\begin{abstract}
Code search is a common practice for developers during software implementation.
The challenges of accurate code search mainly lie in the knowledge gap between source code and natural language (i.e., queries). Due to the limited code-query pairs and large code-description pairs available, the prior studies based on deep learning techniques focus on learning the semantic matching relation between source code and corresponding description texts for the task, and hypothesize that the semantic gap between descriptions and user queries is marginal. In this work, we 
found that the code search models trained on code-description pairs may not perform well on user queries, which indicates the semantic distance between queries and code descriptions.
To mitigate the semantic distance 
for more effective code search, we propose \tool, a \textbf{Que}ry-enriched \textbf{Co}de \textbf{s}earch model. \tool learns to generate \textit{semantic enriched queries} to capture the key semantics of given queries with reinforcement learning (RL).
With RL, the code search performance is considered as a reward for producing accurate semantic enriched queries. The enriched queries are finally employed for code search. Experiments on the benchmark datasets show that \tool can significantly outperform the state-of-the-art code search models.
\end{abstract}




\maketitle

\section{Introduction}\label{sec:Introduction}
\begin{figure}[t]
\centering
\includegraphics[width=0.42\textwidth]{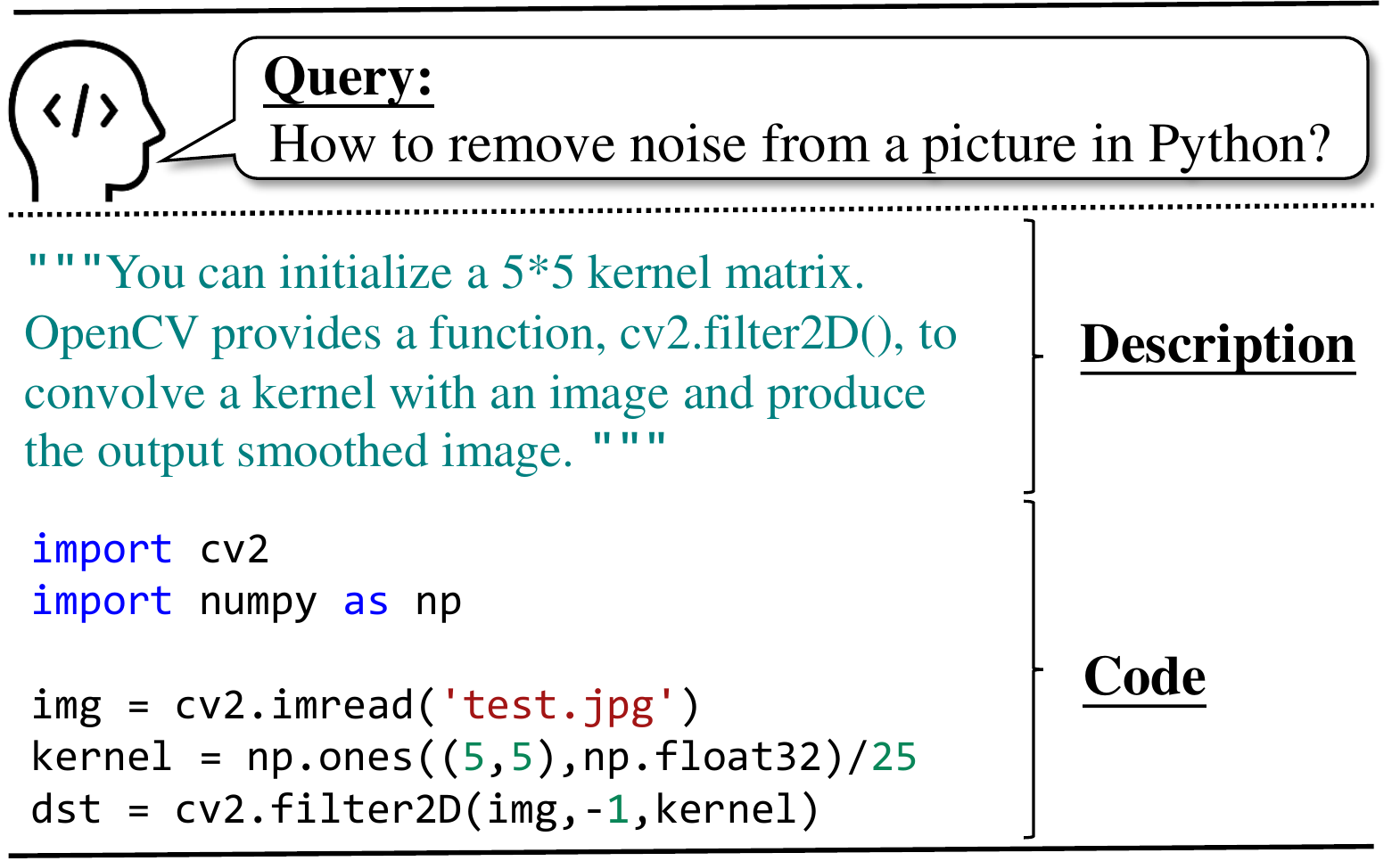}
\caption{An example of a triple of query, description and code.}
\label{fig:pair_example}
\end{figure}

Searching large corpus of existing source code is a common behavior for developers during software programming. The goal of code search is to retrieve code snippets that most closely match a developer's query, which is generally described in natural language (e.g., the query illustrated in the top of Figure~\ref{fig:pair_example}). Existing code search approaches can be divided into two categories: information retrieval (IR)-based (e.g., \cite{DBLP:journals/datamine/LinsteadBNRLB09,DBLP:conf/icse/McMillanGPXF11,DBLP:conf/kbse/LvZLWZZ15}) and deep learning (DL)-based (e.g., \cite{DBLP:conf/pldi/SachdevLLKS018,DBLP:conf/sigsoft/CambroneroLKS019,DBLP:conf/kbse/WanSSXZ0Y19,DBLP:conf/www/YaoPS19}). For example, Linstead et al. develop Sourcerer which retrieves similar code snippets based on software textual content and structural information~\cite{DBLP:journals/datamine/LinsteadBNRLB09}. The IR-based approaches rely on the overlapping tokens or language structures between natural language texts and code snippets, thus suffering from mismatches between the two heterogeneous sources~\cite{DBLP:conf/icse/McMillanGPXF11}. Recent studies resort to deep learning techniques to remedy the issue by embedding source code and textual code descriptions into the same semantic space. The descriptions are usually written by developers to depict the functions of  code snippets (as shown in Figure~\ref{fig:pair_example}), and can facilitate program comprehension. For example, Zhu et al. propose OCoR to capture the character-level and word-level overlaps between code and descriptions based on convolutional network and self attention mechanism~\cite{zhu2020ocor}.




Although promising results were achieved by the deep learning-based approaches, the approaches mostly focused on learning the semantic matching relations between source code and descriptions, and ignored the knowledge gap between input queries and code descriptions. As shown in Figure~\ref{fig:pair_example}, the code search models trained on pairs of code and descriptions can retrieve the code snippet given the description, since there exist semantically-related keywords between the code and description such as ``\textit{cv2}'' and ``\textit{filter2D}''. However, given the user query - ``\textit{How to remove noise from a picture in Python?}'', which is relatively shorter than the description and limited in context, it will be difficult for the trained models to accurately retrieve the corresponding code snippet.
To illustrate the semantic gap between queries and descriptions, in this work, we crawled 11,252 Java and 26,237 Python-related code-description-query triples from GitHub and Stack Overflow, one popular question-answering site for developers, respectively. As far as we know, we are the first to prepare such dataset. We found that the lengths of queries are generally shorter than those of the corresponding code descriptions, \textit{e.g.}, 11.97 v.s. 73.61 words on average in Python (as depicted in Table~\ref{tab:dataset} in Section~\ref{sec:Experimental}). 
Preliminary experimentation further indicates that employing code-description pairs as training dataset possibly do not generalize well for user queries, which motivates us to enrich queries for effective code search.


There also exists prior work on query expansion to enrich the semantic information of user queries~\cite{DBLP:journals/tsc/NieJRSL16,liu2019neural,DBLP:journals/spe/HuangYC19}. For example, Liu et al. predict sets of keywords to expand extremely short queries (e.g., queries with two or three words)~\cite{liu2019neural}. The approaches heavily rely on the relevancy of the extended words to the queries and do not explicitly consider the code search performance during expanding the queries. In this paper, we propose to naturally use the search performance as a reward for capturing the query semantics through reinforcement learning (RL). The proposed approach is named as \textbf{\tool}, an abbreviation of \textbf{Que}ry-enriched \textbf{Co}de \textbf{s}earch model. Specifically, \tool learns to generate the corresponding code descriptions given user queries, and the generated descriptions (called \textit{semantic enriched queries} in the paper) are utilized for improving the code search performance.




In summary, we make the following contributions.
\begin{itemize}

\item We prepare the first dataset containing triples of code, description texts and corresponding search queries. Based on the dataset, we observe that the code search models training on code-description pairs may not perform well on user queries.

\item We propose \tool from a novel perspective of generating semantic enriched queries for code search. \tool naturally captures the key semantic information of queries with the search performance as a reward. The semantic enriched queries are finally utilized to improve the code search task.
\item Extensive experiments demonstrate the effectiveness and flexibility of the proposed model on benchmark datasets. The source code and collected datasets are publicly available through this link\footnote{Hidden url link}.


\end{itemize}

\textbf{Paper structure.} The remainder of the paper is organized as follows. Section~\ref{sec:Methodology} presents our proposed framework.
We introduce the experimental datasets, evaluation metrics and baselines in Section~\ref{sec:Experimental}, and elaborate on the comparison results in Section~\ref{sec:exper}. Section~\ref{sec:Related Work} illustrates the related work. We conclude and mention future work in Section~\ref{sec:conclusion}.


\section{Methodology}\label{sec:Methodology}
\begin{figure*}[t]
\centering
\includegraphics[width=0.7\textwidth]{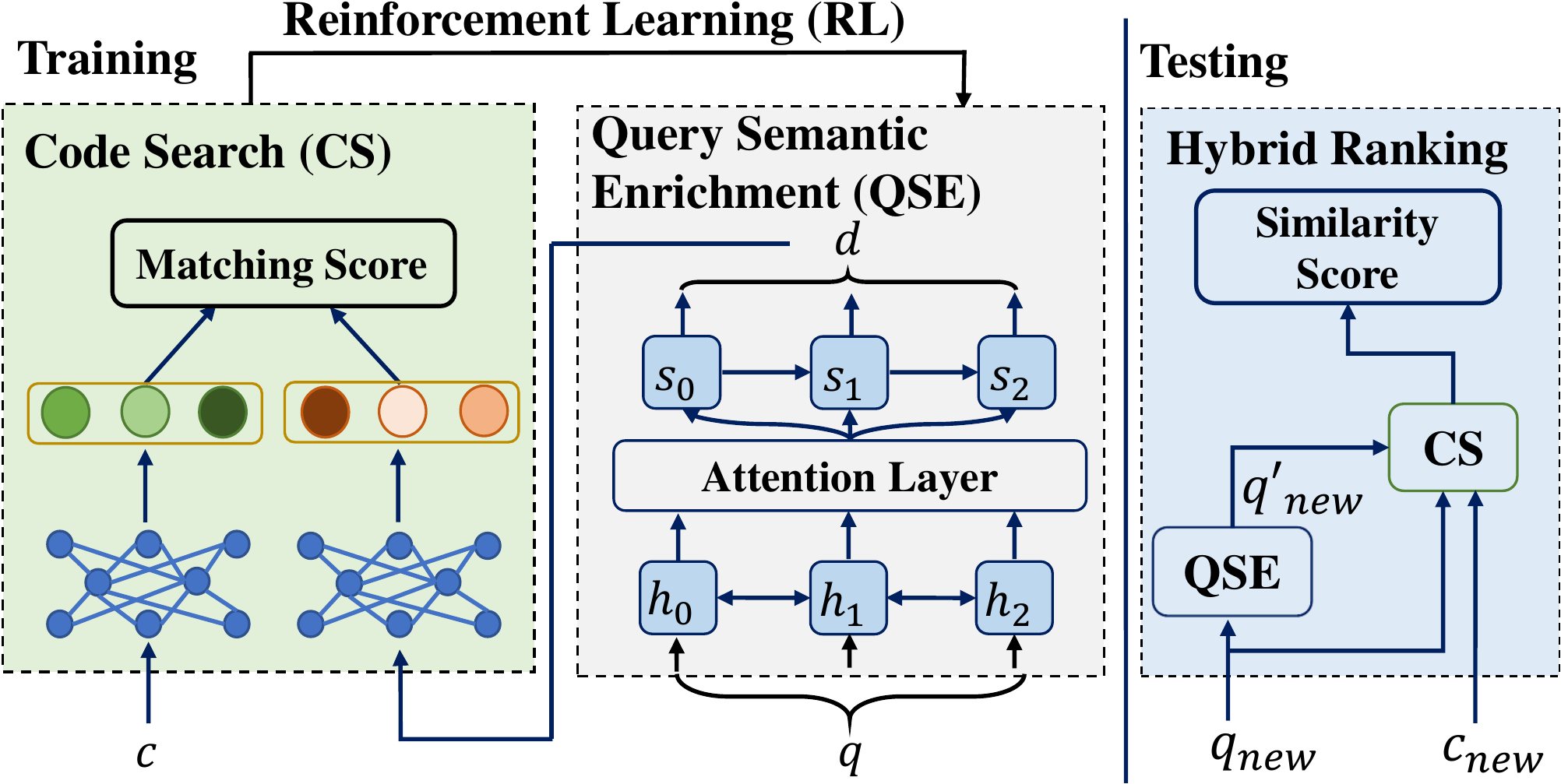}
\caption{Overall framework of the proposed \tool.}
\label{fig:framework}
\end{figure*}

The overview of the proposed approach \tool is depicted in Figure~\ref{fig:framework}. 
As can be seen,
\tool mainly includes three components, i.e., code search model, query semantics enrichment model, and hybrid ranking. Initially, a code search model is trained using large code-description pairs (Section~\ref{subsec:cs}). Then, a query semantic enriching model is designed to generate the corresponding descriptions given user queries based on our collected dataset (Section~\ref{subsec:qd}), during which RL is adopted to enable the code snippets retrieved by the generated descriptions to be ranked higher. The generated descriptions are 
treated as semantic enriched queries and not necessary to be exactly close to the descriptions in the ground truth. Finally, both the semantic enriched queries and original queries are employed for the ultimate code search (Section~\ref{subsec:ranking}). We elaborate on the details of each component in the following.

\subsection{Code Search (CS)}~\label{subsec:cs}
The code search component aims at learning a unified vector representation of both code snippets and descriptions. Since the focus of the work is to mitigate the semantic gap between natural language queries and descriptions for code search, we directly adopt the common framework of the existing code search models~\cite{DBLP:conf/icse/GuZ018,DBLP:conf/sigsoft/CambroneroLKS019}. Let $\langle C,D\rangle$ be the set of code-description pairs, where $C$ and $D$ denote the sets of code snippets and descriptions, respectively. The pipeline of existing code search frameworks~\cite{DBLP:conf/icse/GuZ018,DBLP:journals/corr/abs-1909-09436,zhu2020ocor} are shown in the leftmost part of Figure \ref{fig:framework}, highlighted in green background. Generally, two neural networks are designed separately to encode code and descriptions for mapping the vector representations of the two input sources into the same space, where the neural networks can be Long Short-Term Memory (LSTM), Convolutional Neural Network (CNN) or Transformer~\cite{DBLP:journals/corr/abs-1909-09436}. 

The CS model is trained by minimizing a ranking loss. Specifically, for each description $d\in D$ in the training corpus, the triple of $\langle d, c\text{+}, c\text{-}\rangle$ is prepared as training instance, where $c\text{+}\in C$ and $c\text{-}\in C$ denote the correct code snippet that matches $d$ and sampled negative code snippet that does not match the description (which is randomly sampled from the corpus $C$), respectively. The objective ranking loss function is defined as:

\begin{equation}
   \mathcal{L}(\theta)_{CS}=\sum_{<D,C+,C->}\max(0, \epsilon- sim(\mathbf{d},\mathbf{c}\text{+}) + sim(\mathbf{d},\mathbf{c}\text{-})),
\end{equation}

\noindent where $\theta$ denotes the parameters in the CS model, $\epsilon$ is a constant margin ($\epsilon$ is fixed as 0.05 in all the experiments similar to~\cite{DBLP:conf/icse/GuZ018}), $\mathbf{d}$ denotes the description vector of the description $d$, and $\mathbf{c}\text{+}$ and $\mathbf{c}\text{-}$ denote the code vectors of $c\text{+}$ and $c\text{-}$, respectively. The $sim$ indicates the similarity measurement method which varies for different code search models. 




\subsection{Query Semantic Enrichment (QSE)}\label{subsec:qd}
The query semantic enrichment (QSE) in \tool targets at generating semantic-augmented queries given input queries. The model is trained to generate corresponding human-provided descriptions based on the queries. Unlike existing sequence-to-sequence models~\cite{DBLP:conf/www/YaoPS19}, 
the generated descriptions (i.e., enriched queries) in \tool are not necessary to be exactly close to the human-provided ones, and the purpose of the training is to capture the main semantics of the queries. 

\subsubsection{QSE model}
Given a natural language query $q\in Q$, where $Q$ is the set of collected queries, the QSE learns to produce the corresponding description tokens $d=(d_1,...,d_{N})$.

\begin{equation}
\begin{split}
    p(\mathbf{d}|\mathbf{q}) = p(d_1|d_0, \mathbf{q})\prod_{t=2}^{N} p(d_i| d_{1...t-1}, \mathbf{q})
\end{split}
\end{equation}

We adopt standard bi-directional Long Short-Term Memory (LSTM) model with the attention mechanism~\cite{DBLP:journals/corr/BahdanauCB14} as QSE model structure to generate descriptions.

\begin{equation}
\begin{split}
&h_t = \text{Att-Bi-LSTM}(\overrightarrow{h_{t-1}}, \overleftarrow{h_{t+1}}, q_t), \\
p&(d_t|d_{0...t-1}, \mathbf{q}) = \operatorname{softmax}(\mathbf{W} h_t + b)
\end{split}
\end{equation}

\noindent where $h_t$ is the decoder hidden state at step $t$, $\mathbf{W}$ and $b$ are learnable weights to project the hidden state $h_t$ to the description vocabulary space.

\subsubsection{Training QSE via RL}\label{subsec:rl}
The reinforcement learning (RL) component is designed to render the generated queries in the QSE component 
able to rank the relevant code snippets ahead while close to the semantics of the descriptions. 
We view the description generation process as a Markov Decision Process (MDP)~\cite{bellman1957markovian}, and then use \textit{advantage Actor-Critic} or $A2C$ algorithm~\cite{DBLP:conf/icml/MnihBMGLHSK16} to update the policy gradients.
The MDP mainly consists of four ingredients, including state, action, reward and policy.

\textbf{State:} In the decoding process, a state maintains the input query $\mathbf{q}$ and the generated tokens $d_{1...t-1}$. We take the hidden state vector $s_t$ as the vector representation of the state at the $t$-th decoding step.  

\textbf{Action:} The action for the QSE is to choose the next word $d_t$ from the pre-defined vocabulary. So the action space in our formulation is the vocabulary.

\textbf{Reward:} The reward is used to evaluate the quality of the generated descriptions. In this work, the QSE is rewarded based on whether the correct code snippets retrieved by the semantic enriched queries are ranked higher and the relevancy between the semantic enriched queries and the code descriptions in the ground truth. Hence, we define the reward function $r(s_t, d_t)$ at each time step $t$ as:

\begin{equation}\label{equ:reward}
r(s_t, d_t) =
  \begin{cases}
    \begin{aligned}
    &\alpha*\operatorname{Rank}(C,d_{1...t-1})+ \\
    &\,(1-\alpha)*\operatorname{BLEU}(d_{1...t-1},\mathbf{d}^g),
    \end{aligned}
    & \text{if} \, d_t =<\text{EOS}> \\
    0 &\text{otherwise}
  \end{cases}
\end{equation}

\noindent where $\operatorname{Rank}$ is defined as the popular ranking metric Mean Reciprocal Rank~\cite{DBLP:conf/icse/GuZ018} (detailed in Section~\ref{sec:Experimental}) value, which measures the ranking scores of the correct code snippets given the generated texts. The $\operatorname{BLEU}$~\cite{DBLP:journals/corr/BahdanauCB14} score is the common metric for quantitatively estimating the relevance between generated texts $d_{1...t-1}$ and the descriptions $\mathbf{d}^g$ in the ground truth. We use the BLEU-4 score in the paper.
The $\alpha$ is the reward tunning parameter and $\alpha\in [0,1]$.

\textbf{Policy:} The policy function computes the probability of choosing $d_i$ as the next token. In this work, we use the policy gradient method~\cite{sutton1998introduction} to optimize the policy function.

\subsubsection{Optimization}
In order to stabilize the training process, we resort to the A2C algorithm~\cite{DBLP:conf/icml/MnihBMGLHSK16}. Specifically, the gradient function is defined as:

\begin{equation}\label{equ:a2c}
\nabla\mathcal{L}(\phi)=\mathbb{E}[\sum_{t=1}^{N}(R_t(s_t, d_t) - V(s_t))\nabla\log P(d_t|d_{1...t-1}, C; \phi)],
\end{equation}

\noindent where $R_t(s_t, d_t)=\sum_{t'\ge t}r(s_{t'}, d_{t'})$ is the return for generating word $d_t$ given state $s_t$, $V(s_t)$ is the state value function that estimates the future reward given $s_t$, we  construct a critic network to predict $V(s_t)$ and optimize it with a mean square error (MSE) loss $\mathcal{L}(\rho) = \mathbb{E}_{D\sim P(\cdot|Q)}[\sum^{|D|}_{t=1}(V(s_t;\rho) - R(Q,D))^2]$.

\subsection{Hybrid Ranking}\label{subsec:ranking}
In the training phase, we first train the code search component and then utilize the RL component to generate semantic enriched queries for the code search task. In the testing phase, we design a hybrid ranking component to consider both the original queries and generated enriched queries for returning the ultimate search results.

Specifically, for each input query $q$, we derive the semantic enriched queries $q'$ via the QSE component. The final similarity matching score between the query $q$ and candidate code snippet $c$ is computed as:

\begin{equation}\label{equ:hranking}
    score(\mathbf{q},\mathbf{c}) = \beta*sim(\mathbf{q}\mathbf{'}, \mathbf{c}) + (1-\beta)*sim(\mathbf{q}, \mathbf{c}),
\end{equation}

\noindent where $\beta$ is a parameter to be adjusted experimentally, $0\leq\beta\leq1$. $\mathbf{q}$, and $\mathbf{q'}$, and $\mathbf{c}$ denote the encoded vectors of $q$, $q'$, and $c$, respectively.
\section{Experimental Setup} \label{sec:Experimental}

\subsection{Dataset}

\begin{table*}[t]
\centering
\caption{Statistics of the CodeSearchNet dataset and our collected dataset. Our collected dataset contains aligned code, description texts, and queries. The statistics include the number of pairs and the average lengths of code snippets, descriptions and queries. "-" means there are no such data in the dataset.}\label{tab:codesearchnet}
\scalebox{1.0}{
\begin{tabular}{c|rrrrrr}
\toprule
\textbf{Dataset} & \begin{tabular}{@{}c@{}}\textbf{Training} \\ \textbf{Set}\end{tabular}  & \begin{tabular}{@{}c@{}}\textbf{Validation} \\ \textbf{Set}\end{tabular} & \begin{tabular}{@{}c@{}}\textbf{Test} \\ \textbf{Set}\end{tabular} &
\begin{tabular}{@{}c@{}}\textbf{Avg. Code} \\ \textbf{Token Len.}\end{tabular} &
\begin{tabular}{@{}c@{}}\textbf{Avg. Desc.} \\ \textbf{Token Len.}\end{tabular} &
\begin{tabular}{@{}c@{}}\textbf{Avg. Query} \\ \textbf{Token Len.}\end{tabular}\\
\midrule
\begin{tabular}{@{}c@{}}Java \\ (CodeSearchNet)\end{tabular} & 450,941& 15,053& 26,717 & 162.41 & 38.87 & -\\
\midrule
\begin{tabular}{@{}c@{}}Java \\ (Ours)\end{tabular} & 9,015& 1,117& 1,120 & 506.34 & 55.75 & 12.91 \\
\midrule
\midrule
\begin{tabular}{@{}c@{}}Python \\ (CodeSearchNet)\end{tabular} & 409,230& 22,906& 22,104 & 167.51 & 54.21 & -\\
\midrule
\begin{tabular}{@{}c@{}}Python \\ (Ours)\end{tabular} & 21,009& 2,596& 2,632 & 131.21 & 73.61 & 11.97\\

\bottomrule
\end{tabular}\label{tab:dataset}
}

\end{table*}

\subsubsection{CodeSearchNet.}
CodeSearchNet~\cite{DBLP:journals/corr/abs-1909-09436} is a publicly-available GitHub repository. It provides thousands of pairs of code snippets and the corresponding natural language descriptions. We conducted preprocessing following Gu's work~\cite{DBLP:conf/icse/GuZ018} by first splitting source code tokens of the form CamelCase and snake case to respective sub-tokens, and then taking lowercase. We removed empty items afterwards. The numbers of the subject code-description pairs for the Java and Python datasets are shown in Table \ref{tab:codesearchnet}. 

\subsubsection{Our Collected Datasets.}
Since no dataset containing aligned code snippets, descriptions, and queries were released, we decided to prepare such dataset. The dataset was crawled based on the SOTorrent dataset~\cite{DBLP:conf/msr/BaltesDT008}. SOTorrent comprises pairs of SO (Stack Overflow) posts and GitHub files, where the GitHub files have referred to the corresponding posts, i.e., the code descriptions in the GitHub files clearly cite the post URLs\footnote{{Developers write the SO post URLs in the code descriptions for future program comprehension.}}. Since we focus on the code written in Java and Python, we filtered the code snippets in other languages out. In SOTorrent, only links of the SO posts and GitHub files were provided, so we designed a crawler to traverse all the GitHub links, locate and save the code snippets referred to the SO posts and the corresponding code descriptions. To capture the SO queries, we accessed the SO posts via the SO links and obtained the queries according to the HTML element. The preprocessing procedure for the crawled data is also similar to \cite{DBLP:conf/icse/GuZ018}'s work. We finally split the preprocessed data to be training set, valid set and test set by 8:1:1. The statistics of the subject data are illustrated in Table~\ref{tab:codesearchnet}.

\subsection{Evaluation Metric}\label{tab:metric}
Following~\cite{DBLP:conf/icse/GuZ018,DBLP:journals/corr/abs-1909-09436,zhu2020ocor}, we evaluate the model performance with the standard metrics, including R@k and MRR (Mean Reciprocal Rank). 

\subsubsection{R@k} 
$R@k$ is a common metric to evaluate whether an approach can retrieve the correct answer in the top $k$ returning results. It is widely used by many studies on the code search task. The metric is calculated as follows:
\begin{equation}
    R@k = \frac{1}{|Q|}\sum^{|Q|}_{q=1}\delta(FRank_q\leq k),
\end{equation}
\noindent where $Q$ denotes the query set and $FRank_q$ denotes the rank of the correct answer for query $q$. The function $\delta(Frank_q\leq k)$ $\in \{0, 1\}$ returns 1 if the rank of the correct answer within the top $k$ returning results otherwise returns 0. A higher $R@k$ indicates a better code search performance.
\subsubsection{MRR}
Mean Reciprocal Rank (MRR) is the average of the reciprocal ranks of the correct answers of query set $Q$ ordered by matching score., which is another popular evaluation metric for the code search task. The metric MRR is calculated as follows:

\begin{equation}
    MRR = \frac{1}{|Q|}\sum^{|Q|}_{q=1}\frac{1}{FRank_q}.
\end{equation}

The higher the MRR value is, the better capacity the model has.







\subsection{Baselines}\label{subsec:baseline}
Since the code search component in \tool can be built upon any existing code search models, we mainly choose two popular approaches and one state-of-the-art approach for evaluation. \textbf{\underline{DeepCS}}~\cite{DBLP:conf/icse/GuZ018} is one popular code search model, which employs neural networks such as RNNs to embed both code snippets and descriptions into a joint vector space. Besides the code tokens, it also considers API sequences and method names for learning the representations of code snippets. Since the API sequences are hard to be extracted for the experimental datasets, we slightly modify the original model to only combine method names and code tokens for learning the code representations. \textbf{\underline{UNIF}}~\cite{DBLP:conf/sigsoft/CambroneroLKS019} is similar to DeepCS but with attention mechanism involved. Besides, UNIF only considers code tokens during embedding code snippets. \textbf{\underline{OCoR}}~\cite{zhu2020ocor} is one of the state-of-the-art models. It captures the overlaps between code and descriptions in two levels, including character level and word/identifier level, for calculating the similarities between code and descriptions.

We also choose one popular query expansion approach, denoted as \textbf{\underline{QE}}~\cite{DBLP:conf/wcre/LuSWLD15}, as baseline. QE utilizes WordNet~\cite{leacock1998combining} to extend the queries with synonyms for effective code search. It can also be built upon any code search models. 






\subsection{Implementation Details}
Following the prior studies~\cite{DBLP:conf/icse/GuZ018,DBLP:journals/corr/abs-1909-09436,zhu2020ocor}, we tokenized code snippets, descriptions, and queries. Specifically, code snippets are split in camel case and snake case.  Then we involved the top 10,000 words in the training set as the vocabularies of code snippets, descriptions, and queries, respectively, according to the word frequencies. In addition, we used two symbols to represents the beginning and end of a sentence for sentence generation in  QSE. The word embeddings in both CS and QSE models are randomly initialized. The numbers of hidden states in the encoder and decoder are both defined as 256. The parameter $\alpha$ (in Equ. \ref{equ:reward}) to tune the reward function is set as 1.0, and the parameter $\beta$ (in Equ.~\ref{equ:hranking}) for adjusting the hybrid ranking score is defined as 0.6. Detailed analysis about the parameter settings will be introduced in Section~\ref{sub:parameter_analysis}. We use the Adam optimizer with a learning rate 1e-3 and the learning rate decay ratio is defined as 0.5. For the baseline models, we use the same model hyper-parameters as the original papers. For evaluation, we fix a set of 999 negative code snippets $c\text{-}$ for each test pair $\langle q,c\text{+}\rangle$ following~\cite{DBLP:journals/corr/abs-1909-09436}.

During the training phase, we first train the CS model for 120 epochs, and then train the QSE model for another 20 epochs. Subsequently, we start to train the reinforcement learning component by pretraining the critic network for 10 epochs and jointly training the critic and actor networks for 40 epochs. All the experiments run on a server with 8 * Nvidia Tesla P100 and each one has 16GB graphic memory. Depending on the involved CS model in \tool, the training process lasts from 70 to 240 hours.

\section{Results}\label{sec:exper}
Based on the preceding experimental setup, we first illustrate the semantic gap between queries and descriptions, and then evaluate the effectiveness of the proposed \tool model from three aspects, including comparison with the baselines, parameter analysis, and case study.

\begin{table*}[t]
\centering
\caption{Performance of the baseline models using different evaluation sets.}\label{tab:eval_query}
\scalebox{1.0}{
\begin{tabular}{l|c|rrrr||rrrr}
\toprule
\multirow{2}{*}{\textbf{Approach}} & \textbf{Evaluation} &\multicolumn{4}{c||}{\textbf{Java}} & \multicolumn{4}{c}{\textbf{Python}} \\

& \textbf{Set} & \textbf{R@1} & \textbf{R@5} & \textbf{R@10} & \textbf{MRR} & \textbf{R@1} & \textbf{R@5} & \textbf{R@10} & \textbf{MRR} \\
\midrule
\multirow{2}{*}{\textbf{DeepCS}} & Description & 0.202& 0.389& 0.488& 0.297& 0.160& 0.298& 0.368& 0.231\\
& Query & 0.180& 0.381& 0.483& 0.278& 0.100& 0.223& 0.287& 0.161\\
\midrule
\multirow{2}{*}{\textbf{UNIF}} & Description & 0.213& 0.450& 0.561& 0.324& 0.130& 0.276& 0.347& 0.202\\
& Query & 0.156& 0.439& 0.568& 0.285& 0.078& 0.202& 0.261& 0.143\\
\midrule
\multirow{2}{*}{\textbf{OCoR}} & Description & 0.292& 0.561& 0.654& 0.412& 0.230& 0.423& 0.507& 0.322\\
& Query & 0.211& 0.460& 0.583& 0.330& 0.112& 0.282& 0.363& 0.196\\
\bottomrule
\end{tabular}
}

\end{table*}

\subsection{Preliminary Experiments}\label{subsec:pre}
Here we investigate whether there exists semantic gap between user queries and code descriptions. Specifically, we analyzed the performance of the model trained with code-description pairs on different types of evaluation sets, including queries and descriptions. We choose the three baseline models, i.e., DeepCS, UNIF, and OCoR for experimentation. All the models are trained on the code-description pairs provided by CodeSearchNet~\cite{DBLP:journals/corr/abs-1909-09436}, and then evaluated with the crawled query test set and description test set in Table~\ref{tab:dataset}. The results are illustrated in the Table~\ref{tab:eval_query}. We can observe that using queries as test set presents lower performance than using descriptions as test set for all the three code search baselines. For example, the MRR values decrease by 12.8\% and 32.9\% on average on the Java and Python datasets, respectively; and the R@1 values also show a downtrend, with the average decreasing rates at 21.8\% and 42.9\% on the two datasets, respectively. The results imply the semantic gap between queries and description texts. Thus, models trained on code-description pairs possibly do not perform well for practical user queries, and mitigating the semantic gap between queries and descriptions is necessary for more accurate code search.







\begin{table*}[t]
\centering
\caption{Comparison results with baseline models. The bold figures indicate the best results.}\label{tab:results}
\scalebox{1.0}{
\begin{tabular}{l|rrrr||rrrr}
\toprule
\multirow{2}{*}{\textbf{Approach}} & \multicolumn{4}{c||}{\textbf{Java}} & \multicolumn{4}{c}{\textbf{Python}} \\

& \textbf{R@1} & \textbf{R@5} & \textbf{R@10} & \textbf{MRR} & \textbf{R@1} & \textbf{R@5} & \textbf{R@10} & \textbf{MRR} \\
\midrule
\midrule
\multicolumn{9}{c}{\textbf{DeepCS-based}}\\
\midrule
\textbf{DeepCS} & 0.180& 0.381& 0.483& 0.278 & 0.100& 0.223& 0.287& 0.161 \\
\textbf{QE}& 0.169& 0.361& 0.464& 0.266 & 0.105& 0.222& 0.288& 0.167\\
\textbf{QueCos} w/o RL & 0.110& 0.239& 0.322& 0.180& 0.048& 0.131& 0.180& 0.095\\
\textbf{QueCos} w/o HR & 0.192 & 0.402 & 0.498 & 0.294 & 0.117 &0.252&0.309	& 0.182
\\
\textbf{QueCos} (ours)& \textbf{0.282}& \textbf{0.519}& \textbf{0.611}& \textbf{0.394}  & \textbf{0.149}& \textbf{0.297}& \textbf{0.370}& \textbf{0.224} \\
\midrule
\midrule
\multicolumn{9}{c}{\textbf{UNIF-based}}\\
\midrule
\textbf{UNIF}  & 0.156& 0.439& 0.568& 0.285 & 0.078& 0.202& 0.261& 0.143\\
\textbf{QE}& 0.155& 0.431& 0.553& 0.281 & 0.072& 0.179& 0.250& 0.133\\
\textbf{QueCos} w/o RL & 0.098& 0.296& 0.408& 0.197& 0.066& 0.168& 0.231& 0.122\\
\textbf{QueCos} w/o HR & 0.159&0.450 & 0.565 &0.293 & \textbf{0.128} & 0.293 & 0.359 & 0.207\\
\textbf{QueCos} (ours)& \textbf{0.199}& \textbf{0.517}& \textbf{0.660}& \textbf{0.347} & 0.125& \textbf{0.319}& \textbf{0.394}& \textbf{0.217} \\
\midrule
\midrule
\multicolumn{9}{c}{\textbf{OCoR-based}}\\
\midrule
\textbf{OCoR} & 0.211& 0.460& 0.583& 0.330 & 0.112& 0.282& 0.363& 0.196\\
\textbf{QE}& 0.227& 0.511& 0.647& 0.358& 0.115& 0.291& 0.380& 0.202\\
\textbf{QueCos} w/o RL & 0.183& 0.418& 0.504& 0.289& 0.095& 0.227& 0.305& 0.163\\
\textbf{QueCos} w/o HR & 0.216 & 0.516 & 0.613 & 0.349 & \textbf{0.184} & \textbf{0.408} & \textbf{0.492} & 0.282\\
\textbf{QueCos} (ours)& \textbf{0.242}& \textbf{0.553}& \textbf{0.649}& \textbf{0.375} & \textbf{0.184}& 0.398& 0.489& \textbf{0.289}\\
\bottomrule
\end{tabular}
}

\end{table*}

\subsection{Main Comparison Results}
We compare \tool with the four baseline models introduced in Section~\ref{sec:Experimental} on the collected datasets. Since the code search component in \tool and QE can be built upon any code search models, we analyze the performance with each of the code search baselines involved. The comparison results are illustrated in Table~\ref{tab:results}. The approach ``\tool w/o RL'' indicates the proposed \tool model without the RL procedure, that is, the code search performance is not considered as reward for the query enrichment. The approach ``\tool w/o HR'' denotes the \tool without the HR component, i.e., the semantic enriched queries generated in QSE component are directly used for code search. Based on the comparison results, we achieve the following observations:


\textbf{Observation 1: \tool significantly improves the baseline models given user queries.} As can be seen in Table~\ref{tab:results}, \tool presents better performance than the corresponding code search model on both Java and Python datasets. For example, \tool increases the accuracy of the corresponding code search model by 33.0\%, 24.7\%, 23.8\% and 21.9\% in terms of R@1, R@5, R@10 and MRR, respectively, for the Java dataset on average.
The results demonstrate that with the query semantics augmented, \tool can more accurately recommend relevant code snippets for given user queries. It is unsurprising that \tool with OCoR achieves the best results on the Python dataset since OCoR already presents the best performance among the three baselines. But we also observe that \tool with DeepCS shows the best results on the Java dataset in terms of R@1 and MRR, which further highlights the effectiveness of \tool for enriching user queries for the task.


\textbf{Observation 2: The query semantic enriching component is more effective than the popular query expansion approach.} By comparing \tool with the popular query expansion approach QE, we can discover that \tool can significantly improve the performance of QE.
However, with QE involved, the code search models may perform worse, for example, QE with DeepCS degrades the performance of DeepCS by 6.11\% in terms of R@1 on the Java dataset. This indicates that explicitly complementing the queries with synonyms may not contribute the model performance. The results also prove the usefulness of the query semantic enriching component in \tool for code search. 

\textbf{Observation 3: The RL and HR components in \tool plays a key role in enriching the query semantics.} Comparing \tool w/o RL with \tool, we can find that \tool performs worse without the RL component, presenting even poorer performance than the corresponding base code search model. The phenomenon indicates that simply using the generated descriptions may bias the semantics of the queries and lead to inaccurate code search. This is reasonable since the queries are generally shorter than the descriptions according to Table~\ref{tab:dataset}, and the model is hard to accurately capture the semantic relations between queries and descriptions through merely training on query-description pairs. We provide an error case in Section \ref{sub:Error Study} for further illustration.

Comparing \tool w/o HR with \tool, we can observe that the effectiveness of only using the generated enriched queries for code search is limited. Integrating the original queries could further enhance the model performance.

\subsection{Parameter Analysis}\label{sub:parameter_analysis}

\begin{figure*}[t]
\centering
\includegraphics[width=0.9\textwidth]{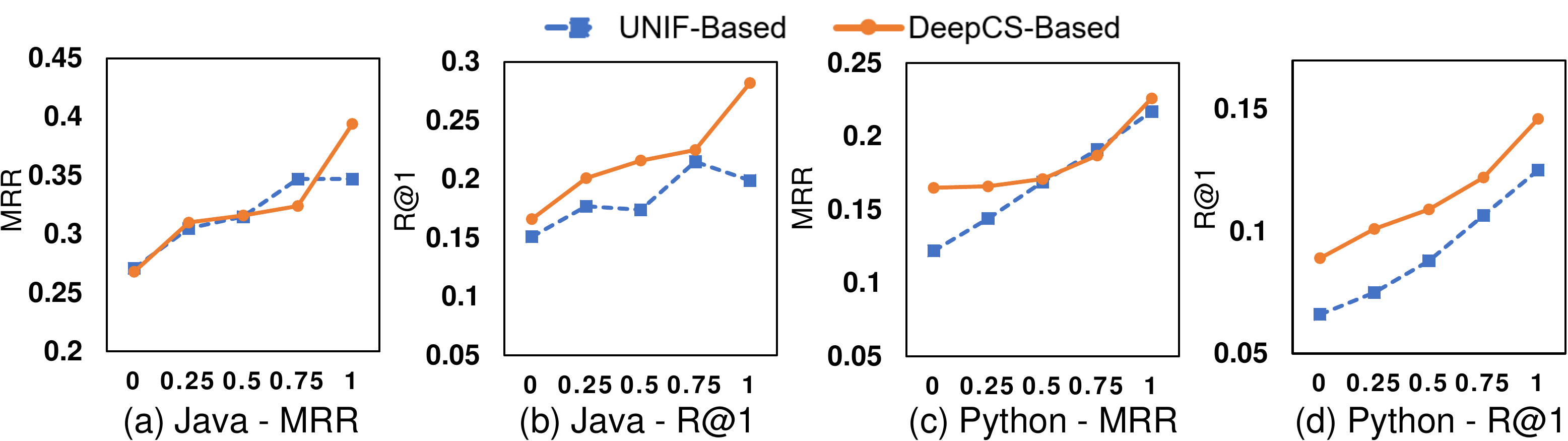}
\caption{Parameter sensitive study on the reward tuning parameter $\alpha$. The parameter analysis for OCoR-based \tool was ignored due to the huge computation resources of OCoR.}
\label{fig:alpha}
\end{figure*}

\begin{figure*}[t]
\centering
\includegraphics[width=0.9\textwidth]{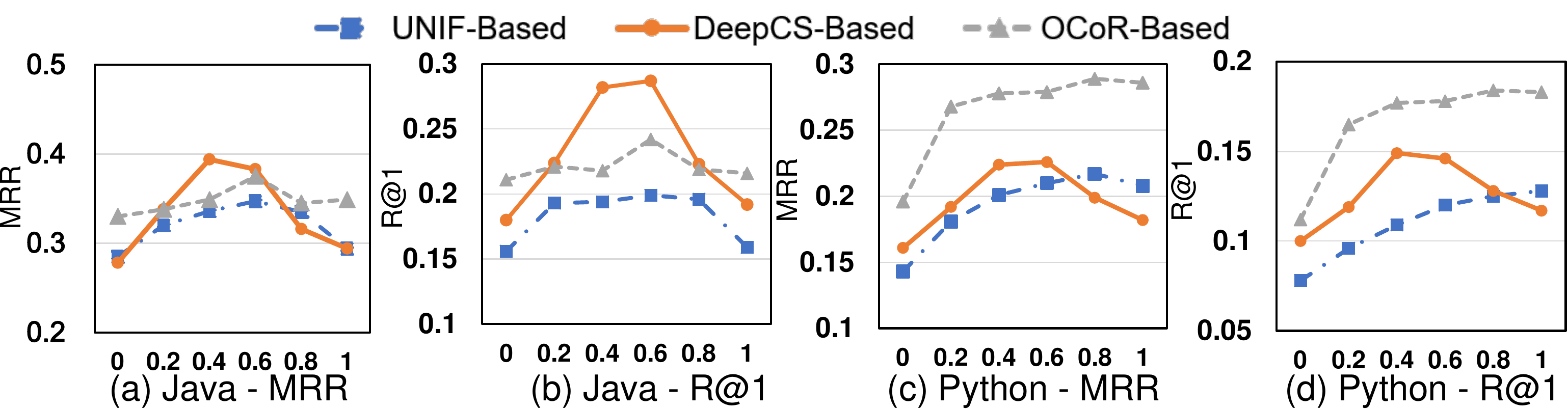}
\caption{Parameter sensitive study on the hybrid ranking parameter $\beta$.}
\label{fig:beta}
\end{figure*}

In the section, we analyze the impact of two important hyper-parameters, including the reward tuning parameter $\alpha$ and the hybrid ranking parameter $\beta$, on the performance of \tool. Figure~\ref{fig:alpha} and Figure~\ref{fig:beta} illustrate the performance variations of \tool with different base code search models as the parameters changes, respectively. During analyzing the impact of $\alpha$, we did not conduct the parameter analysis for the OCoR-based \tool due to the huge computation resources required by OCoR\footnote{The training process of OCoR lasts around 240 hours.}. As can be seen in Figure~\ref{fig:alpha}, when $\alpha$ increases, the model performance presents an obvious uptrend in terms of both R@1 and MRR. The results indicate that considering the similarity between the generated texts and the descriptions in the ground truth as a reward during semantic enriched query generation is unhelpful for the code search task. This is reasonable since only involving the code ranking performance as a reward can render the RL component focus on generating the descriptions that could rank relevant code snippets higher. Thus, we define the reward tuning parameter $\alpha$ as 1.0 during the experiments, i.e., the relevancy between the generated text and the human-provided description is not considered as a reward.

According to Figure~\ref{fig:beta}, 
The performance variations of DeepCS-based \tool along with the value are close to an inverted ``U'' shape, but the performance variations of UNIF-based or OCoR-based \tool present an increasing trend along with the value. From Table 1, we observe that the code descriptions in the Python dataset are relatively longer than those in the Java dataset. We guess the generated enriched queries could deliver more accurate semantics for the Python dataset, so the relevance between the enriched queries and code snippets is more important for the search results comparing to the Java dataset. We fix the value of $\beta$ as 0.6 in the experiments since the model would approach the optimal accuracy. 

\subsection{Case Study}\label{sub:Case Study}

\begin{mintedbox}[fontsize=\scriptsize]{java}
private Boolean equalLists(List<TariffSpecification> listA, List<TariffSpecification> listB) {
	return listA.size() == listB.size() && listA.containsAll(listB);
}
\end{mintedbox}

\begin{table}
\centering
\scalebox{0.96}{\begin{tabular}{rp{5.5cm}}
\toprule
Query & Java compare two lists' object values?\\  
\midrule 
\multirow{2}{*}{\begin{tabular}{@{}c@{}}Description in\\ Ground Truth\end{tabular}} & NOTE: will not work with duplicate objects\\
\midrule
\multirow{4}{*}{\begin{tabular}{@{}c@{}}Query Enriched \\ by QE \end{tabular}} & comparison compare equivalence liken equate two 2 II deuce ii list listing tilt inclination lean name heel number object... \\
\midrule
\multirow{3}{*}{\begin{tabular}{@{}c@{}}Query Enriched \\ by \tool \end{tabular}} & converts list length of code taken its returned list code list\textgreater\, sorts a list using widths as a list integers\\
\bottomrule
\end{tabular}
}
\caption{User query, description in the ground truth, and the enhanced query by QE and \tool for the successful case 1.}\label{tab:case1}
\end{table}

\begin{lstlisting}[frame=none,caption={Successful case 1, with more details listed in Table~\ref{tab:case1}.},captionpos=b,label=lst:case1]
\end{lstlisting}

\begin{mintedbox}[fontsize=\scriptsize]{python}
def _add_subelem(root_element, name, value): 
    if value is None: 
        return
    if type(value) is dict: 
        if name == "link": 
            ET.SubElement(root_element, name, value) 
        elif name == 'content': 
            e = ET.Element(name, type = value['type']) 
            e.append(CDATA(value['content'])) 
            root_element.append(e)
        else:
            subElem = ET.SubElement(root_element, name) 
            for key in value: 
                _add_subelem(subElem, key, value[key])
    else:
        ET.SubElement(root_element, name).text = value
\end{mintedbox}
\begin{lstlisting}[frame=none,caption={Successful case 2, with more details shown in Table~\ref{tab:case2}.},captionpos=b,label=lst:case2]
\end{lstlisting}

Code Listing~\ref{lst:case1} and Table~\ref{tab:case1} show a case in which \tool with UNIF-based successfully returns the correct code snippet of the query ``Java compare two lists' object values?'' at the top of the ranking list, and the basic UNIF model only ranks the code snippet at 11. We can observe that the code description does not indeed reflect the functionality of the code, i.e., comparison of two lists' objective values. The generated semantic enriched query of \tool is ``converts list length of code taken its returned list code list\textgreater\, sorts a list using widths as a list integers'', which could deliver more semantics than the input query although the grammar is not exactly correct. The expanded query by the QE approach is ``comparison compare equivalence liken equate two 2 II deuce ii list listing...'', which also only ranks the code snippet at 11. In addition, \tool with DeepCS-based and OCoR-based can rank the correct code snippet at 9 and 19, respectively, while the basic DeepCS and OCoR provide the rank at 8 and 30 respectively. 
Overall, the generated semantic enriched query can be helpful for more accurate code search.

\begin{table}[h]
\centering
\scalebox{0.96}{\begin{tabular}{rp{5.5cm}}
\toprule
\multirow{2}{*}{Query} & How to output Cdata using element tree?\\  
\midrule 
\multirow{ 4}{*}{\begin{tabular}{@{}c@{}}Description \\ in Ground Truth\end{tabular}} & HORRIBLE HACK! A wee hack too, the content node must be converted to a CDATA block. This is a sort of cheat, see: \\
\midrule
\multirow{ 4}{*}{\begin{tabular}{@{}c@{}}Query Enriched \\ by QE \end{tabular}} & end\_product output yield output\_signal production outturn turnout exploitation victimization victimisation use utilize utilis...\\
\midrule
\multirow{ 4}{*}{\begin{tabular}{@{}c@{}}Query Enriched \\ by \tool \end{tabular}} & xml xml element note retaining elements are quite source extended xml element xml etree node element as value save a tree object from xml to xml map back xml xml\\
\bottomrule
\end{tabular}
}
\caption{User query, description in the ground truth, and the enhanced query by QE and \tool for the successful case 2.}\label{tab:case2}
\end{table}


For the example shown in the Code Listing~\ref{lst:case2} and Table~\ref{tab:case2}, the \tool with UNIF-based improves the rank of the correct code snippet from 60 to 2 compared to the basic UNIF approach~\cite{DBLP:conf/sigsoft/CambroneroLKS019}. The proposed model can capture that the terms ``Cdata'' and ``element tree'' in the query are relevant to the XML language, and the generated semantic enriched query clearly includes terms such as ``xml'', ``element'', and  ``etree''; while the extended query by the QE approach fails to catch the relation of the query to XML, returning the code snippet at 16. So we suppose that through enriching the query semantics, \tool can more accurately rank the code snippets.
\subsection{Error Study}\label{sub:Error Study}


\begin{mintedbox}[fontsize=\scriptsize]{java}
SuppressWarnings(""unchecked"")
Map<String, String> myMap = (Map<String, String>) deserializeMap();
\end{mintedbox}
\begin{lstlisting}[frame=none,caption={An error Case, with more details in Table~\ref{tab:case3}.},captionpos=b,label=lst:case3]
\end{lstlisting}

Code Listing~\ref{lst:case3} and Table~\ref{tab:case3} show an error case. UNIF ranks corresponding code snippet at 4 according to the original query, however, UNIF-based \tool only ranks it as 10. The poor performance can be attributed to that \tool misunderstands the meaning of the word ``\textit{address}'' in the query, and hence the enriched query contains irrelevant keywords, e.g., ``\textit{host}'', ``\textit{url}'', ``\textit{ip}'', etc. In the future, we will design a quality assurance filtering component to distinguish the semantically-irreverent keywords in the generated enriched queries.


\begin{table}
\centering
\scalebox{0.96}{\begin{tabular}{rp{5.5cm}}
\toprule
\multirow{2}{*}{Query} & How do I address unchecked cast warnings?\\  
\midrule 
\begin{tabular}{@{}c@{}}Description \\ in Ground Truth\end{tabular} & None \\
\midrule
\multirow{4}{*}{\begin{tabular}{@{}c@{}}Query Enriched \\ by QE \end{tabular}} & address speech destination savoir-faire turn\_to speak direct call cover accost unbridled cast mold mould form plaster\_cast casting roll...\\
\midrule
\multirow{4}{*}{\begin{tabular}{@{}c@{}}Query Enriched \\ by \tool \end{tabular}} & verify the device class wifi on host ( s , ip listener , url . , it will dynamically handle web fixed allowing the passing and run of the context ( such , domain, dynamic ip address.\\
\bottomrule
\end{tabular}
}
\caption{User query, description in the ground truth, and the enhanced query by QE and \tool for the error case in Code Listing~\ref{lst:case3}.}\label{tab:case3}
\end{table}

\section{Related Work} \label{sec:Related Work}


\subsection{Code Search}
As the rapid growing of open-source code corpus, to retrieve the code fragments satisfying a user's intent with high accuracy concerning developing productivity. 
Prior studies on code search focus on figuring out the latent relationship between NL queries and code snippets.
The work~\cite{DBLP:conf/icse/McMillanGPXF11} proposes Portfolio, modelling the navigation behavior of programmers with random surfer and returning a chain of functions.
Lv et al.~\cite{DBLP:conf/kbse/LvZLWZZ15} propose CodeHow, a code search tool that first retrieves all possible API calls to augment the queries. 

Recently, an increasing number of works using neural networks for code search have been proposed~\cite{liu2019neural,DBLP:conf/www/YaoPS19,DBLP:conf/sigsoft/CambroneroLKS019,zhu2020ocor}. 
Sachdev et al. present a neural network-based model, which creates continuous vector representations of codes for comparison~\cite{DBLP:conf/pldi/SachdevLLKS018}.
In the work \cite{liu2019neural}, Liu et al. present NQE, which predicts the keywords in the NL queries and expands them in a productive way to improve performance for shorter queries. 
Gu et al. propose DeepCS which embeds both the code snippets and queries into a joint vector space to measure the similarity between them~\cite{DBLP:conf/icse/GuZ018}. 
\cite{DBLP:conf/sigsoft/CambroneroLKS019} introduces UNIF, a bag-of-words-based network which converts code snippets and docstring tokens into embedding matrices with supervised learning method.
The authors in~\cite{DBLP:conf/www/YaoPS19} treat code annotation and code search as two divided tasks and use the generated code annotations to improve the performance of code search. However, no previous studies have explicitly involved the search performance for query semantic enrichment for improving the task.
Zhu et al.~\cite{zhu2020ocor} propose an overlap-aware OCoR, which explicitly considers the overlap between code and descriptions in both word and character levels. Based on the word-level and character-level representations of the code and descriptions, OCoR utilizes MLP to compute the relevance.


\subsection{Code Representation Learning}
Code representation learning aims at learning the semantics of programs for facilitating various downstream tasks related to program comprehension, such as code clone detection, code summarization, bug detection~\cite{zhang2019novel,DBLP:conf/icse/GuZ018,DBLP:journals/corr/abs-1909-09436,wang2020modular, leclair2020improved,wan2018improving,li2019improving, lam2017bug}, etc. The development of deep learning techniques boosts the research on code representation learning. In the work~\cite{DBLP:conf/icse/GuZ018,DBLP:journals/corr/abs-1909-09436}, code snippets are split into tokens and fed into neural networks such as RNNs and multi-head attentions for the representation learning. Considering the structural nature of code, \cite{wang2020modular, leclair2020improved, zhang2019novel} combine the abstract syntax trees (ASTs) into neural networks for capturing the code semantics. LeClair et al.~\cite{leclair2020improved} use GNN-based encoder to model the AST of each program subroutine.

Recent work adapts pre-training techniques in the natural language processing field~\cite{devlin2018bert, lan2019albert,liu2019roberta} for better program comprehension. For example, the work~\cite{DBLP:journals/corr/abs-2002-08155} presents CodeBERT, the first bimodal pre-trained model for programming language (PL) and natural language (NL). Guo et al.~\cite{DBLP:journals/corr/abs-2009-08366} later propose GraphCodeBERT, which combines both code tokens and the data flow information during pre-training, and achieves more promising performance.


\section{Conclusions}\label{sec:conclusion}

In this paper, we have proposed a novel RL-based query-enriched code search model, named \tool, for more effective code search. \tool can capture the main semantics of user queries through learning to generate the corresponding descriptions, where RL is adopted to render the generated descriptions to be able to rank the relevant code snippets higher. We also crawled the first dataset containing triples of code, descriptions, and queries. Experiments demonstrate that \tool is more effective than the baseline models and also flexible to incorporate any code search models. In the future, we will combine external knowledge graph and adapt Transformer to further enrich the semantics of the queries.







\printcredits

\bibliographystyle{cas-model2-names}

\bibliography{cas-refs}



\end{document}